\documentclass{optica-article}
\journal{oe}
\articletype{Research Article}
\usepackage{lineno}
\usepackage{algorithm}
\usepackage{algpseudocode}
\usepackage{mathtools}

\begin{document}

\title{Experimental demonstration of a Grover-Michelson interferometer}

\author{Christopher R. Schwarze,\authormark{1, *} David S. Simon,\authormark{1,2}, Anthony D. Manni,\authormark{1} Abdoulaye Ndao,\authormark{1,3} and Alexander V. Sergienko\authormark{1,4}}
\address{\authormark{1}Department of Electrical and Computer Engineering \& Photonics Center, Boston University, 8 Saint Mary’s St., Boston, Massachusetts 02215, USA\\
\authormark{2}Department of Physics and Astronomy, Stonehill College, 320 Washington Street, Easton, Massachusetts 02357, USA\\
\authormark{3}Department of Electrical and Computer Engineering, University of California, San Diego, La Jolla, California 92093-0401, USA\\
\authormark{4}Department of Physics, Boston University, 590 Commonwealth Avenue, Boston, Massachusetts 02215, USA}
\email{\authormark{*}crs2@bu.edu}

\begin{abstract}
We present a low-resource and robust optical implementation of the four-dimensional Grover coin, a four-port linear-optical scatterer that augments the low dimensionality of a regular beam-splitter. While prior realizations of the Grover coin required a potentially unstable ring-cavity to be formed, this version of the scatterer does not exhibit any internal interference. When this Grover coin is placed in another system, it can be used for interferometry with a higher-dimensional set of optical field modes. In this case, we formed a Grover-Michelson interferometer, which results when the traditional beam-splitter of a Michelson interferometer is replaced with a four-port Grover coin. This replacement has been shown to remove a phase parameter redundancy in the original Michelson system, now allowing continuous tuning of the shape and slope of the interference pattern. We observed an intensity interferogram with $97\%$ visibility and a phase sensitivity more than an order of magnitude larger than a regular Michelson interferometer. Because this device is readily formed with nearly the same number of optomechanical resources as a Michelson interferometer, but can outperform it drastically in phase delay evaluation, it has a great potential to improve many interferometric sensing and control systems.
\end{abstract}
\section{Introduction\label{intro}}
Interferometry has long been a leading application of electrodynamics. It is not only responsible for several core systems for high-precision metrology and control, but in addition, optical interferometers also form a key component in more complex optical systems due to their ability to manipulate light in both a highly extensive and predictable manner \cite{BerkhoutKoenderink, Bogaerts2020, Pérez-López2020, mi13040614, Poon:04}. For example, optical interferometers represent the physical implementation of tunable photonic gates, making them a key component of many optical computing platforms \cite{RevModPhys.79.135}. 

Nonetheless, virtually all modern interferometric systems heavily rely on a standard tried and true methodology: using common, low-dimensional optical scatterers to separate and combine electromagnetic energy in a well-known arrangement, such as the Michelson, Mach-Zehnder, Sagnac, or Fabry-Perot configuration. A property common to all of these systems is that the scattering events they admit can only connect two input and two output field modes at a time. Assuming the scattering transformations are linear and lossless, this property implies the underlying scattering transformations can always be represented by a $2\times 2$ matrix. 

\begin{figure}[ht]
    \centering
    \includegraphics[width=.7\textwidth]{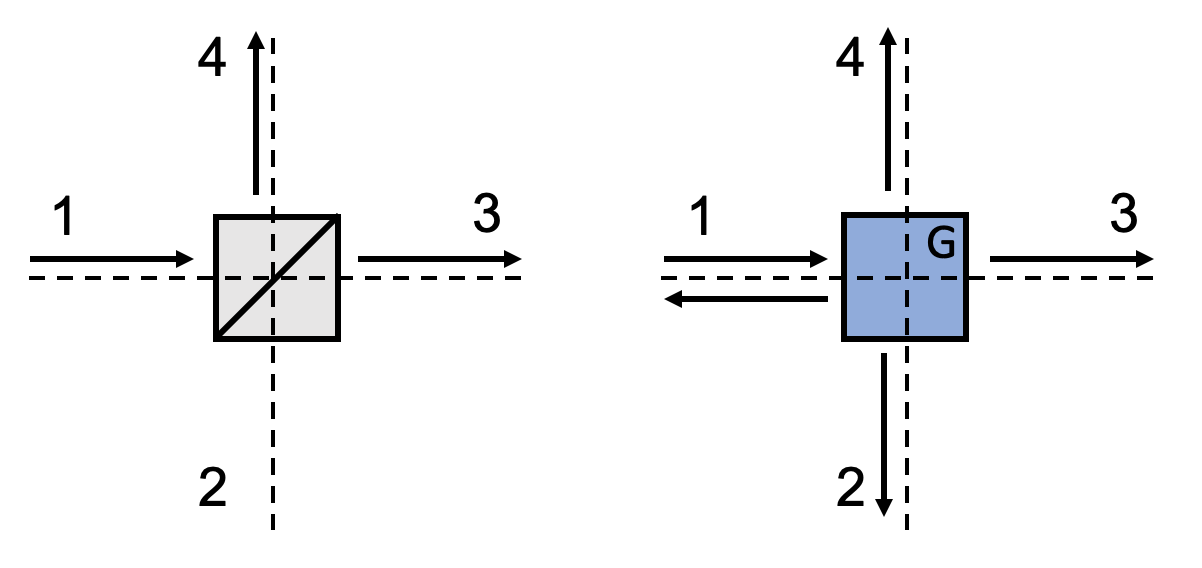}
    \caption{(left) Optical beam-splitter. Light entering any given port can only emerge from two output ports. The ports excited for input to port 1 are indicated by the black arrows. This restriction causes this four-port device to act as a two-dimensional scatterer. (right) Grover four-port, a higher-dimensional generalization to the beam-splitter. This device equally excites four ports instead of two for each input. Because this four-dimensional scatterer has four-ports, it can replace the beam-splitter in any optical system, which has been shown to carry benefits in well-known interferometric configurations.}
    \label{fig:bsg}
\end{figure}

This low-dimensional structure is to be expected for the partially-reflective plates that form a Fabry-Perot etalon, since (at normal incidence) these surfaces can only be accessed from two input/output ports, one on either side. On the other hand, this $2\times 2$ restriction represents a stark deficiency in the beam-splitter that is universally regarded as a fundamental component in optical systems, such as the other interferometers named above. Whether it be a tabletop cube formed from two connected prisms, a glass plate with a thin metallic or dielectric coating, a fiber coupler, or one of the many variants found on photonic chips, \textit{a beam-splitter is a four-port device that is confined to act in a two-dimensional space}. The property responsible for this restriction is known as \textit{directional-bias}: light entering any given port of the beam-splitter cannot reverse direction. In the left half of Fig. \ref{fig:bsg} a generic beam-splitter is depicted schematically. The black arrows illustrate the directions light can flow for the case of a beam incident on port 1. No light can emerge from ports 1 and 2. No matter which port is used for input, only two ports are excited for output.

When all of the fields coupled by a lossless, 50:50 beam-splitter are placed in a single basis that contains both the input and output modes for all four ports, the device's scattering matrix may be written as
\begin{equation}
B = \frac{1}{\sqrt{2}}
\begin{pmatrix}\label{eq:bs}
    0 & 0 & 1 & 1\\
    0 & 0 & 1 & -1\\
    1 & 1 & 0 & 0\\
    1 & -1 & 0 & 0\\
\end{pmatrix}.
\end{equation}
Here, the input and output modes for a given port are identified with one another. This formalism is often harmless but in cases where ambiguities arise due to counter-propagating excitations in the same spatial (and polarization) mode, the ingoing and outgoing modes may be labeled separately, in which case the true beam-splitter scattering matrix would be $8\times 8$. 

Traditional interferometry has seen great successes in spite of its confinement to low-dimensional scatterers. However, the above discussion begs the following question: how might interferometers change if higher-dimensional scatterers were used? A suitable candidate for this investigation is the four-port Grover coin, 
\begin{equation}\label{eq:grov}
G = \frac12
\begin{pmatrix}
    -1 & 1 & 1 & 1\\
    1 & -1 & 1 & 1\\
    1 & 1 & -1 & 1\\
    1 & 1 & 1 & -1\\
\end{pmatrix}.
\end{equation}
This device is suitable specifically because it also has four input/output ports, so it is able to replace the beam-splitter in any scenario. Moreover, it also has equal splitting at each input, but now, \textit{this amplitude division is equally distributed over four outputs instead of two}. This Grover coin belongs to a more general class of \textit{directionally-unbiased} linear scatterers which are defined by their coherent back-reflections. It is shown schematically next to the beam-splitter in Fig. \ref{fig:bsg}, right.

Replacing the traditional beam-splitter with a Grover four-port in common interferometer configurations such as the Michelson \cite{PhysRevA.107.052615} and Mach-Zehnder \cite{PhysRevA.106.033706} has been shown to provide a variety of benefits. In the Mach-Zehnder configuration, multiple phases can be measured simultaneously in a single system. In the upgraded Michelson configuration, which is demonstrated in this work, the interference pattern can be continuously deformed. This allows continuous control of the output sensitivity, in turn enabling super-resolution phase measurements to be made with a simple configuration. 

Resolution enhancements in an interferometer typically stem from an increase in slope of the output intensity with respect to the input phase. This can result from many techniques, such as using a shorter wavelength or using maximally-entangled photon states of the form $(|N \ 0\rangle + |0\ N\rangle)/\sqrt{2}$, since an $N$-photon bunch will acquire $N$-times the optical phase as a single photon. Resonant interferometers like the Fabry-Perot etalon use multiple passes to acquire a sharper phase response. This is how the Grover-Michelson interferometer functions: because the centerpiece Grover-coin is directionally-unbiased, light impinging the coin from an arm can be redirected back into the same arm or transferred into the other arm. In other words, the single-pass arms of a traditional Michelson interferometer are converted to strongly-coupled optical resonators. This coupling allows the sharpness of the phase response of one arm to be controlled by the other arm, as will be discussed in detail later in this paper. 

Low-dimensional (two-port) directionally-unbiased devices are not uncommon. For example, active (phase-tunable) two-port unbiased devices such as the standard Michelson and Fabry-Perot interferometers have long been built from low-dimensional passive scatterers like mirrors, beam-splitters, and phase shifters. Less common are implementations of higher-dimensional, directionally-unbiased multiports \cite{PhysRevA.93.043845, Osawa:18, Kim:21}, due to the number of resources required or challenge in aligning and/or stabilizing them. 

In this paper, we first report a very stable implementation of the Grover-four port, which reduces the complexity of the previous design \cite{PhysRevA.93.043845, PhysRevA.101.032118} by taking advantage of a special symmetric arrangement of optical components. Not only does the device function without a resonator, but it now no longer interferes any beams if used alone. After introducing this improved design of the four-port Grover coin in the next section, we review the enhanced resolution behavior of the Grover-Michelson interferometer \cite{PhysRevA.107.052615} in Section 3. In Section 4 we present results from the experimental implementation of the Grover coin and the Grover-Michelson interferometer formed from it. Some applications and final conclusions are discussed in Section 5. 

\section{Generalized Grover four-port}

\begin{figure}[ht]
    \centering
    \includegraphics[width=.7\textwidth]{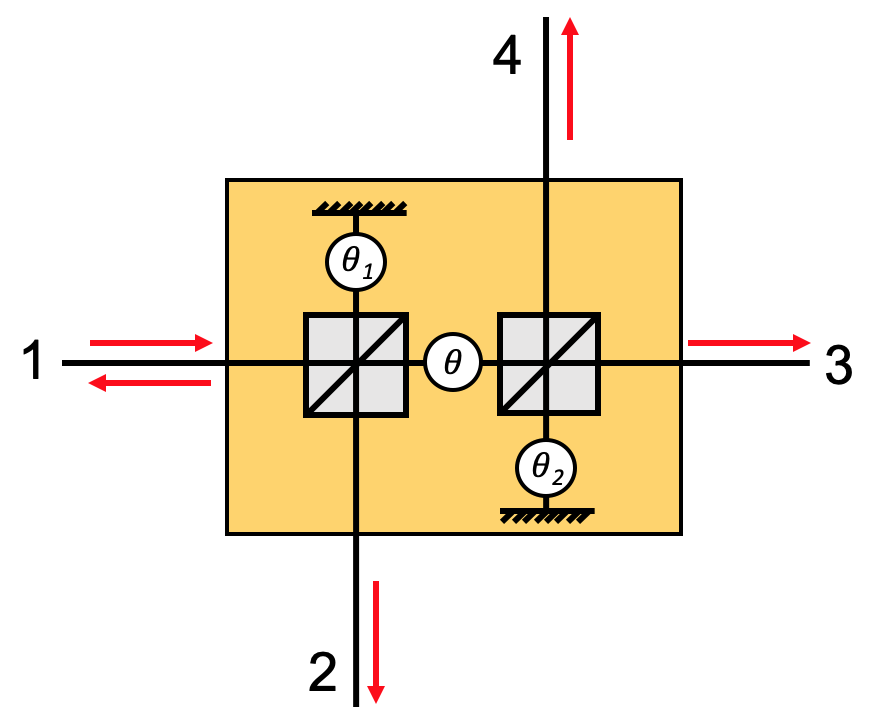}
    \caption{Schematic of an optical Grover four-port. This implementation does not exhibit any internal interference. The beam-splitters are each 50:50 for a single but otherwise arbitrary polarization. When $\theta$ is zero, the output exactly matches equation (\ref{eq:grov}). The phases $\theta_1$ and $\theta_2$ do not matter, as they can be nullified by adding external phase shifts, which cannot be measured.}
    \label{fig:gc}
\end{figure}

Consider the arrangement shown in Fig. \ref{fig:gc}. It is easy to verify that a resonator will not form when the device is used from any input port. In fact, no beams are overlapped, so interference will not occur. We assume the beam-splitters divide energy equally for a single but otherwise arbitrary polarization. Because no interference is occurring, the scattering probabilities are only determined by the number of beam-splitter encounters. This number is easily seen to be two in all input cases, so that the scattering probabilities will all be 1/4.
The scattering matrix for the beam-splitters will be that of equation (\ref{eq:bs}). Other beam-splitter phase conventions can be used without loss of generality as these can be represented with external phase shifts placed around each beam-splitter. This could only result in a global shift of the device parameters  $\theta_1, \theta_2$ and $\theta$, which is experimentally undetectable. Here, the above convention is such that the $\pi$ phase shift is applied to the reflected portion of the output state whenever the input accesses the beam-splitter from the facet coupled to a mirror and the neighboring facet which couples in and out of the device (at the ports labeled 1 and 3 in Fig. \ref{fig:gc}). This convention is used to ensure that the Grover coin will later emerge at a simple value of the parameters $\theta_1, \theta_2$ and $\theta$.

If $\theta_1 = \theta_2$ the system in Fig. \ref{fig:gc} is invariant with respect to 180-degree rotations, so in general, the system will behave identically when input to port 1 as it does when input to port 3 (and vice versa), except that $\theta_1$ and $\theta_2$ are swapped in the output state. The same goes for ports 2 and 4. Hence an analysis of inputs to ports 1 and 2 is sufficient to derive the full scattering matrix. 

To that end, we begin with a photon input to port 1. Half of the optical state transmits to the other beam-splitter with amplitude $e^{i\theta}/\sqrt{2}$ which then splits by the second beam-splitter, giving amplitudes of $e^{i\theta}/2$ at both ports 3 and 4. The other half of the state acquires a round-trip phase of $\theta_1$ inside the mirror-arm. Inside this mirror-arm, there is also a separate phase contribution of $\pi$ each from the beam-splitter reflection and the end-mirror reflection, but when these contributions are combined together, the state is left unchanged. This portion $e^{i\theta_1}/\sqrt{2}$ exiting the mirror-arm then re-enters the beam-splitter and splits, ending with an amplitude of $-e^{i\theta_1}/2$ at port 1 and $e^{i\theta_1}/2$ at port 2. 

The analysis is similar for port 2. The initially reflected portion also crosses to the other side, leaving ports 3 and 4 both with amplitude $e^{i\theta}/2$. The transmitted piece acquires $1/\sqrt{2}$. After striking the mirror while acquiring the round-trip phase $\theta_1$ this amplitude is $-e^{i\theta_1}/\sqrt{2}$. Therefore when transmitting through the beam-splitter a second time and emerging at port 2 the amplitude is $-e^{i\theta_1}/2$. Exiting port 1 the amplitude is $e^{i\theta_1}/2$ due to the additional $\pi$ phase shift acquired from the beam-splitter reflection. Placing all outputs into a scattering matrix results in a scattering matrix of
\begin{equation}
    G = \frac{1}{2}
\begin{pmatrix}
    -e^{i\theta_1} & e^{i\theta_1} & e^{i\theta} & e^{i\theta}\\
    e^{i\theta_1} & -e^{i\theta_1} & e^{i\theta} & e^{i\theta}\\
    e^{i\theta} & e^{i\theta} & -e^{i\theta_2} & e^{i\theta_2}\\
    e^{i\theta} & e^{i\theta} & e^{i\theta_2} & -e^{i\theta_2}
\end{pmatrix}.
\end{equation} 
When $\theta_1 = \theta_2 = \theta = 0$ the system is precisely a Grover four-port. In fact, placing external phases outside this device with values $-\theta_1/2$ at ports 1 and 2 and $-\theta_2/2$ at ports 3 and 4 removes $\theta_1$ and $\theta_2$ while only translating $\theta$ by $-(\theta_1 + \theta_2)/2$. Thus for any values of $\theta_1$ and $\theta_2$, $\theta$ can be redefined to account for that global phase translation, and in accord with this, the above scattering matrix can be expressed in the simplified but fully general form
\begin{equation}\label{eq:gengc}
    G = \frac{1}{2}
\begin{pmatrix}
    -1 & 1 & e^{i\theta} & e^{i\theta}\\
    1 & -1 & e^{i\theta} & e^{i\theta}\\
    e^{i\theta} & e^{i\theta} & -1 & 1\\
    e^{i\theta} & e^{i\theta} & 1 & -1
\end{pmatrix}.
\end{equation} 
This scattering action cannot be realized with the passive (parameter-free) Grover coin of eq. (\ref{eq:grov}), even with external phase shifts.

\section{Generalized Grover-Michelson interferometer}
\begin{figure}[ht]
    \centering
    \includegraphics[width=\textwidth]{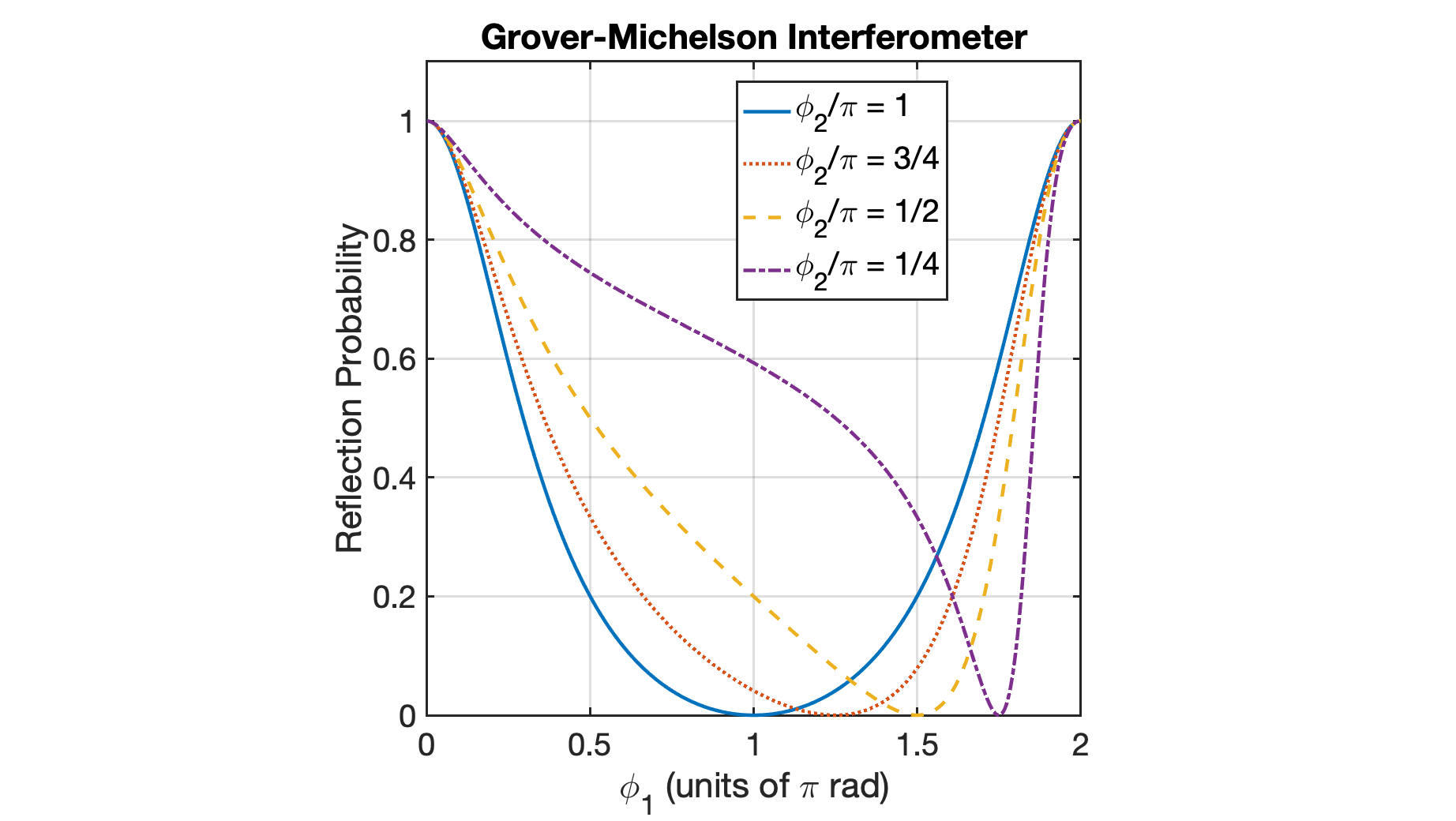}
    \caption{Reflection probability vs. arm phase $\phi_1$ in a Grover-Michelson interferometer. Varying $\phi_2$ leads to deformation of the probability curve, allowing the maximum slope to be increased.}
    \label{fig:theory}
\end{figure}

With the four-port generalized Grover coin of eq. (\ref{eq:gengc}), a Michelson-like interferometer can be formed by placing mirrors at any two of the open ports of the device described above and shown in Fig. \ref{fig:gc} (see Fig. \ref{fig:gmi}). We assume here that this is done at ports 3 and 4, and in those arms the acquired phases are $\phi_1$ and $\phi_2$, respectively. When the bridge phase $\theta = 0$, the output can be found analytically \cite{PhysRevA.107.052615}; a derivation is shown in the appendix of this paper. Defining
\begin{align}
    B &\coloneqq \frac12 (e^{i\phi_1} + e^{i\phi_2}), \\
    C &\coloneqq \frac12 (e^{i\phi_1}-e^{i\phi_2}).
\end{align}
Then for a photon input to port 1, $|\psi_0\rangle = a_1^\dagger|0\rangle$, the output state is given by 
\begin{align}\label{eq:gmi}
|\psi_{\text{out}}\rangle &= \bigg [
\bigg (\frac{C^2}{2B - 2} - \frac{B}{2} - \frac12\bigg )a_1^\dagger + \bigg (\frac{C^2}{2B - 2} - \frac{B}{2} + \frac12\bigg )a_2^\dagger\bigg]|0\rangle \\ \notag &\coloneqq (r(\phi_1, \phi_2) a_1^\dagger + t(\phi_1, \phi_2) a_2^\dagger)|0\rangle.
\end{align}
The reflection probability and transmission probability are respectively given by $R(\phi_1, \phi_2) = |r(\phi_1, \phi_2)|^2, T(\phi_1, \phi_2) = |t(\phi_1, \phi_2)|^2$. Since $C$ only appears in $|\psi_{\text{out}}\rangle$ as $C^2$, this output state is left unchanged when the parameters $\phi_1$ and $\phi_2$ are exchanged. Thus, without loss of generality, we can let $\phi_2$ define a reflection (or transmission) probability curve and $\phi_1$ define a point of interest on this curve. As $\phi_2$ is changed, the curve is continuously deformed. Some examples are shown in Fig. \ref{fig:theory}. Nonzero values of $\theta$ break the permutation symmetry of the Grover matrix, causing the curves produced to change in a way that depends on which ports are sealed with end-mirrors. Thus, $\theta$ varies the shape of the curves seen in a given homotopy and must be stabilized to select a specific curve. However, at values other than zero, $\theta$ does not otherwise diminish the main effect of a tunable maximum slope. The details of this system are beyond the scope of this work and will be described elsewhere.

\section{Experimental results}
\subsection{Implementation of four-port Grover coin}
Using two cube beam-splitters, we first assembled the generalized four-port Grover coin shown in Fig. \ref{fig:gc} and measured the output probability distribution for each input port. A narrowband NECSEL semiconductor laser with wavelength $\lambda = 633 $ nm was used. It was selected for its very long coherence length $\ell_c > 1$ km. Since the internal paths in our device were at most only a few dozen centimeters, this source remains coherent over many round-trips through the device, which is important when the device is later converted to a Grover-Michelson interferometer.

The cube beam-splitters were nominally 50:50 power splitting, but in reality were about $R = 48\%$ and $T = 52\%$, with variations from input to input of a couple percent. The measured probability matrix of the Grover coin was
\begin{equation}\label{eq:p}
P =
\begin{pmatrix}
0.2615&    0.2396   & 0.2228&    0.2390\\
0.2424 &   0.2277  &  0.2434 &   0.2554\\
0.2218  &  0.2428 &   0.2404  &  0.2372\\
0.2377   & 0.2579&    0.2390   & 0.2248
\end{pmatrix}
\pm 
\begin{pmatrix}
0.0008&    0.0003   & 0.0011&    0.0004\\
0.0007 &   0.0009  &  0.0006 &   0.0016\\
0.0012  &  0.0008 &   0.0028  &  0.0011\\
0.0006   & 0.0012&    0.0001   & 0.0016
\end{pmatrix}.
\end{equation}
The diagonal terms corresponding to back-reflections were measured using an additional beam-splitter to separate the input and output light, as depicted in the ``input isolation'' box of Fig. \ref{fig:gmi}. The columns of the above matrix are normalized by the input power and therefore contain information about system losses.

The effects of losses and scattering imbalances can be separated by re-normalizing each column by its sum. This scaling guarantees that the columns have a sum of 1 and a mean value of $\frac14$. Understanding of imbalances is more significant than losses in correlated photon experiments, since in these experiments losses only increase the data collection time. Applying this scaling leads to the following matrix $P_0$:
\begin{equation}\label{eq:p0}
P_0 =
\begin{pmatrix}
0.2715&    0.2475   & 0.2357&    0.2499\\
0.2516 &   0.2353  &  0.2574 &   0.2671\\
0.2302  &  0.2508 &   0.2542  &  0.2480\\
0.2467   & 0.2664&    0.2527   & 0.2350\\
\end{pmatrix}
\pm
\begin{pmatrix}
0.0008&    0.0003   & 0.0012&    0.0004\\
0.0008 &   0.0009  &  0.0006 &   0.0017\\
0.0012  &  0.0008 &   0.0030  &  0.0011\\
0.0006   & 0.0012&    0.0001   & 0.0016
\end{pmatrix}
\end{equation}
The uncertainties in the equations (\ref{eq:p}) and (\ref{eq:p0}) above originate from relative variations in input beam and output detector alignment. Each port was independently probed for input and output three times to obtain those estimates for alignment uncertainty. Fluctuations of the source laser were measured to be an order of magnitude lower than these alignment variations. Regardless, the above alignment variations are still noticeably less than some of the discrepancies between the numerical values measured and those of the ideal Grover four-port probability matrix, where each value is $\frac14$. These discrepancies are primarily inherited deviations of the beam-splitters from their ideal 50:50 values. The beam-splitter imbalances sometimes work with each other and sometimes against each other. Overall, keeping required resources to a minimum is critical in multiport design, since the device's deviations from ideal behavior (both imbalances and losses) will accumulate more drastically as the number of resources increases.

The measurements for the above data gathered nearly all of the light in the cross section of the beam. However, within a free-space interferometer like this, often only the center of the beam is measured; that is the flattest portion of the beam and is accordingly the closest approximation to the theoretical plane wave assumption. This portion of the beam can lose energy to the outer portion of the beam due to diffraction, especially if the beam is passed through small pinholes. Thus, in those setups, the actual losses of a single pass through the system can be noticeably worse, but could be mitigated by using an integrated waveguide platform. 

\subsection{Grover-Michelson interferometer}
\begin{figure}[ht]
    \centering
    \includegraphics[width=.9\textwidth]{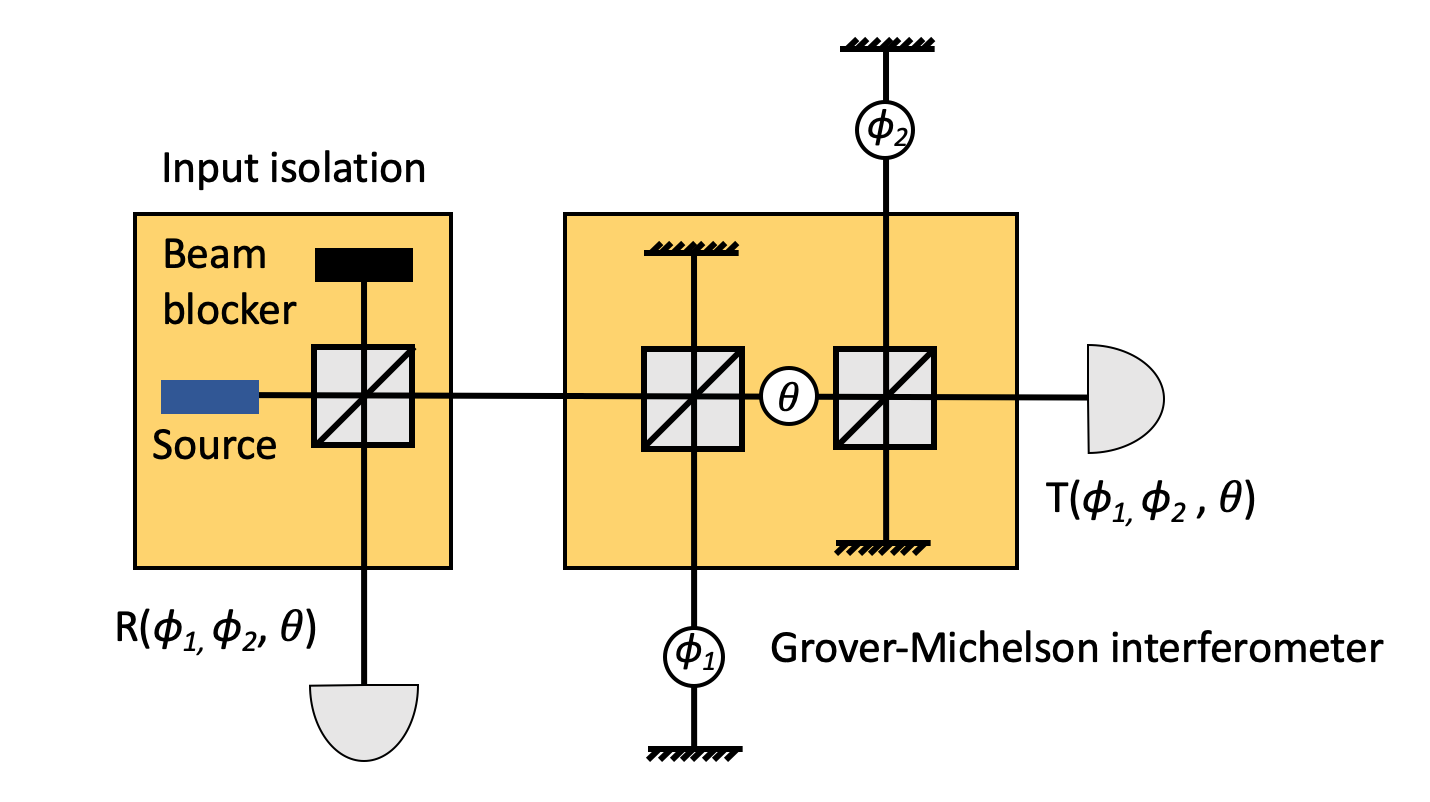}
    \caption{Experimental setup of Grover-Michelson interferometer. There are two main components: the source block (left) and the Grover-Michelson interferometer (right). Placing each block on a separate platform allowed the Grover coin probability matrix entries to be measured for each input. A beam-splitter is placed in front of the source to separate parts of the counter-propagating modes entering and exiting the Grover-Michelson block. A narrowband NECSEL semiconductor laser with wavelength $\lambda = 633$ nm was used. Phases $\phi_1, \phi_2$, and $\theta$ were controlled with piezoelectric actuators. Optical power meters were used to measure the output; a lens and pinhole were used to the pick off the quasi-planar center of the beam before hitting the detector.}
    \label{fig:gmi}
\end{figure}

After placing two external mirrors around the generalized Grover four-port coin, a Grover-Michelson interferometer was obtained. The experimental setup is shown in Fig. \ref{fig:gmi}.
After conducting preliminary measurements of both transmission and reflection curves, it was recognized that the curves for reflection exhibited slightly better visibility than their transmission counterparts. This was assumed to be the result of better internal beam-splitter coefficient balancing for light traveling in those paths between the source and reflection-side detector. The visibility of these curves remained very good even in sensitive regions.

Subsequently, several measurements of reflected intensity $R(\phi_1, \phi_2, \theta)$ were conducted. Data was collected using a Newport 1835-C power meter sampling at 1 kHz. Onboard averaging of $N = 10$ points was conducted. Phase parameters were all controlled with open-loop piezoelectric actuators, mounted to standard optomechanical translation stages. The piezo crystals were held in place by the spring-induced contact force between the stage and the tip of the micrometer which is used to translate the stage.
\begin{figure}[ht]
    \centering
    \includegraphics[width=\textwidth]{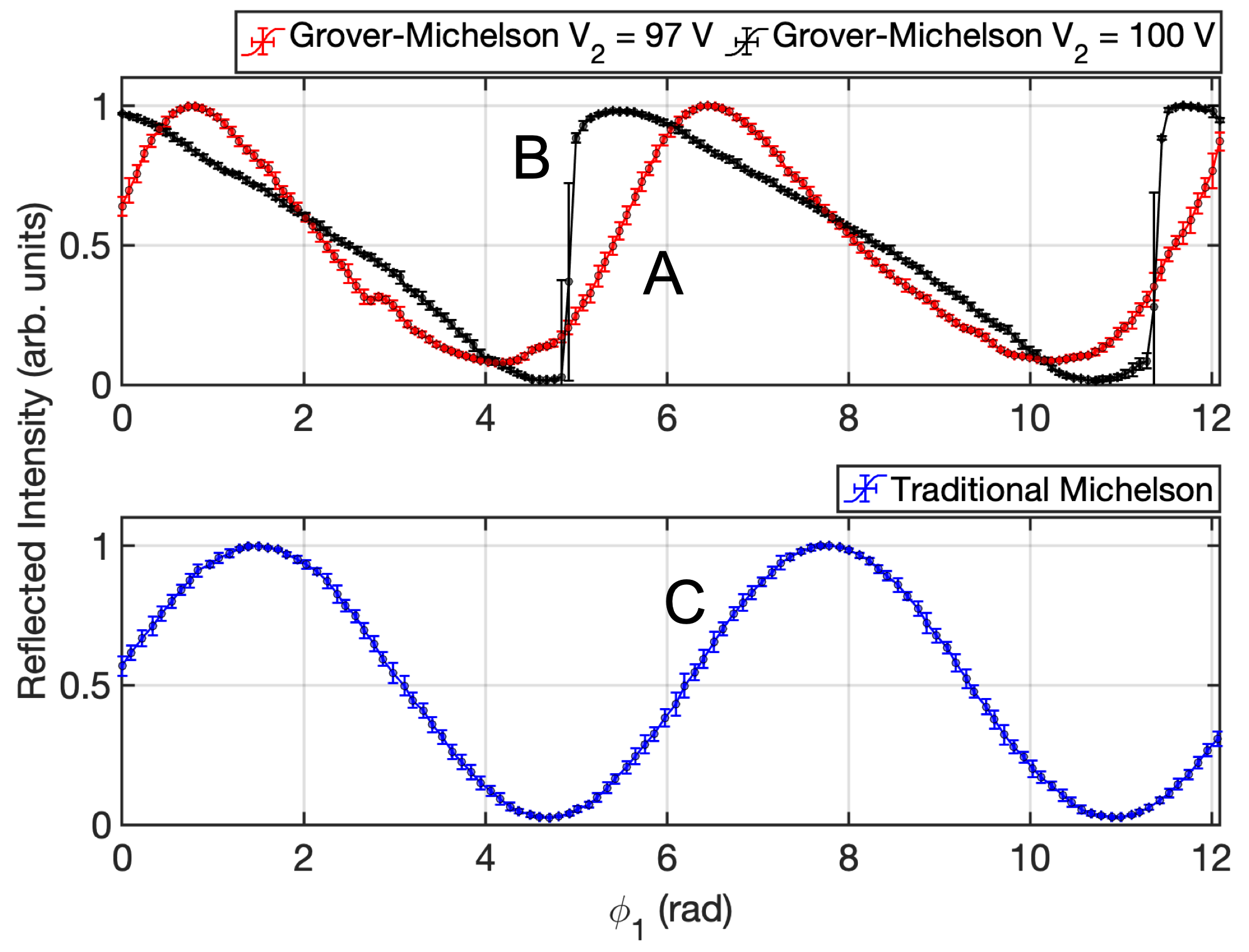}
    \caption{Reflected intensity interferograms of a Grover-Michelson interferometer (top, A \& B) and regular Michelson interferometer (bottom, C) vs. arm phase $\phi_1$. Varying $\phi_2$ through the piezo voltage $V_2$ allowed the more symmetric Grover-Michelson outputs like the red curve (A) to be skewed into the black curve (B). This improves the maximum slope by over an order of magnitude compared to the regular Michelson (C), while maintaining high visibility in these regions. Note fluctuations are also reduced in the extremely linear low-slope regions of the black curve. Errorbars are shown corresponding to one standard deviation of the $N = 10$ point averages of the power measurements. In some regions the errorbar separation is smaller than the point size of the plot. The error in $\phi_1$ is below 2 mrad, which is also too small to be seen.}
    \label{fig:data}
\end{figure}

Two examples within the Grover-Michelson family of curves are compared with a curve collected from a regular Michelson interferometer in Fig. \ref{fig:data}. A more symmetric curve indicating the dependence of the intensity on the phase $\phi_1$ is shown in red. Generally as $\phi_1$ was scanned, $\phi_2$ and $\theta$ were passively stabilized by the floating optical bench and an airflow-insulating enclosure. When the $\phi_2$ phase was changed by increasing the voltage on the second piezo crystal by $\Delta V = 3$ V, the curve as a function of $\phi_1$ became very linear on one side and sharply sloped on the other. This interferogram exhibited a visibility of 97$\%$ and maximum slope greater than $7$, which is over an order of magnitude larger than the $0.5$ theoretical maximum of a Michelson (assuming the double-pass arm phase is called $\phi$ instead of $2\phi$). The very linear, flat sloped region also exhibited an expected decrease in fluctuations. When the curve is altered by changing the piezo voltage $V_2$, the visibility is seen to vary as well. This is expected, as the curve-tuning effectively changes the finesse of the system; this is what allows the phase-response to be increased. However, at the same time, losses in the system will accumulate in different amounts when the finesse changes. The balancing between different internal paths inside the system will change as a consequence, inducing changes in output visibility.

\section{Discussion}

\begin{figure}[ht]
    \centering
    \includegraphics[width=\textwidth]{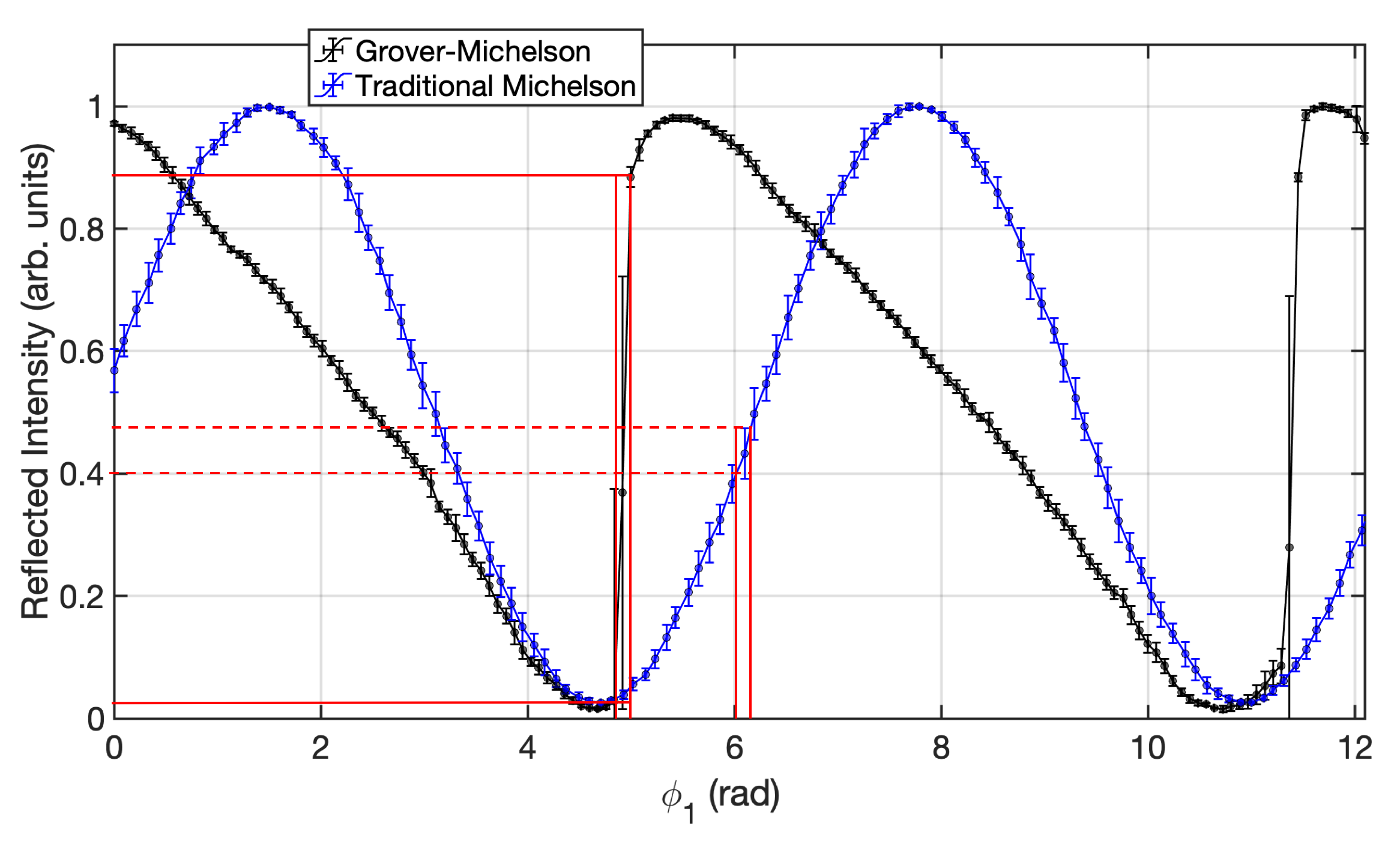}
    \caption{Overlay of Grover-Michelson data (black) and traditional Michelson data (blue). The red lines illustrate how an equal change in phase $\Delta \phi_1$ (vertical line pairs) can generate a much larger response in intensity $\Delta I$ in the Grover-Michelson interferometer (solid horizontal lines) than in a traditional Michelson interferometer (dashed horizontal lines). For this equal value of $\Delta \phi_1$, the intensity change is 12.6 times larger for the Grover-Michelson interferometer.}
    \label{fig:data2}
\end{figure}

In this paper we formed an efficient Grover four-port and demonstrated unbiased multiport interferometry with it. The Grover-Michelson device here can be realized with nearly the same number of optomechanical resources as a traditional Michelson interferometer (or about the same as a Mach-Zehnder). However, due to the new degrees of freedom accessed by the centerpiece Grover coin, the Grover-Michelson is able to significantly outperform its predecessor. For the same change in phase, the Grover-Michelson interferometer can generate a much larger intensity modulation, as shown in Fig. \ref{fig:data2}. The two pairs of vertical red bars are the same width, representing the same change in phase $\Delta \phi_1$, which could be enacted through a change in path-length $\Delta \phi_1(k) = k n (\Delta \ell)$ or a refractive index change inducing $\Delta \phi_1(k) = k (\Delta n) \ell$. The corresponding change in intensity $\Delta I$ is encapsulated by the horizontal red bars (solid for Grover-Michelson, dashed for Michelson) and is roughly 12.6 times larger for the Grover-Michelson device. This device therefore has the potential to enhance the performance of a vast number of interferometric tools, such as super-resolution phase sensors or optical intensity modulators. 

The flat-sloped regions would also be useful, for example, if the device were to act as a tunable beam-splitter in a larger system; this region is extremely linear in $\phi_1$ and exhibits higher dynamic range as well as lower fluctuations. Being able to tune the slope is also useful in applications where too large of a phase shift might be unwanted, since it could bring the system into another period, causing a readout ambiguity. This is a prominent issue in phase contrast microscopy \cite{Kurata:24, pc1, pc2}. Tuning the slope would allow different phase scales to be probed independently, somewhat analogous to a zoom lens. A fast scanning, closed-loop system would be advantageous in the future to allow locking onto a specific curve that could be viewed and altered in real-time.

Being a multi-pass system, the device will exhibit an increasingly narrowband response as the system is tuned to have a higher slope. This feature might have future uses, since different wavelengths will have different slopes, and thus get (de)modulated in different manners. At the same time, such an interferometer could operate with a greater resolution at any specific wavelength by a simple adjustment of the control arm phase. Employing a birefringent phase shift element would allow different polarizations to have different slopes, potentially improving sensing systems which use a polarization quadrature readout, such as homodyne quadrature laser interferometry \cite{HQLI}.

Overall, we have illustrated the beginnings of a very versatile yet simple interferometer. It remains to be seen how this system and potentially other interferometers formed from unbiased multiports will be integrated into new optical technologies.

\section*{Funding.}
Air Force Office of Scientific Research MURI award number FA9550-22-1-0312.
\section*{Disclosures.}
None to report.
\section*{Data availability.}
All data generated for this article is available upon reasonable request to the authors.
\section*{Appendix: Calculation of Grover-Michelson scattering amplitudes}

Consider a Grover coin $G$ given by the matrix in eq. (\ref{eq:grov}). We will seal port 3 with a phase shift $\phi_1$ and mirror and do the same for port 4 with a phase shift $\phi_2$ and mirror. Collectively the linear phase transformations these devices enact during each round trip will be denoted $\Phi$ and $M$. Using the following definitions,
\begin{align*}
    A &\coloneqq \frac12(-a_1^\dagger + a_2^\dagger), \\ 
    B &\coloneqq \frac12 (e^{i\phi_1} + e^{i\phi_2}), \\
    C &\coloneqq \frac12 (e^{i\phi_1}-e^{i\phi_2}),
\end{align*}
we will first consider excitations of $(a_3^\dagger \pm a_4^\dagger)$ undergoing single round-trip in the interferometer arms. Assuming the light is initially propagating away from the Grover coin, we have
\begin{subequations}
\begin{align}
(a_3^\dagger + a_4^\dagger) &\xrightarrow{M, \Phi} -e^{i\phi_1}a_3^\dagger -e^{i\phi_2}a_4^\dagger \\&\xrightarrow{G} -\frac12 (e^{i\phi_1}(a_1^\dagger + a_2^\dagger - a_3^\dagger + a_4^\dagger)+  e^{i\phi_2}(a_1^\dagger + a_2^\dagger + a_3^\dagger - a_4^\dagger)) \\&= -\frac12 (e^{i\phi_1} + e^{i\phi_2})(a_1^\dagger+a_2^\dagger)+ \frac12 (e^{i\phi_1}-e^{i\phi_2})(a_3^\dagger - a_4^\dagger) \\ &= -B(a_1^\dagger+a_2^\dagger) + C(a_3^\dagger - a_4^\dagger)\label{eq:1}
\end{align}
\end{subequations}
while
\begin{subequations}
\begin{align}
(a_3^\dagger - a_4^\dagger) &\xrightarrow{M, \Phi} - e^{i\phi_1}a_3^\dagger + e^{i\phi_2}a_4^\dagger \\ &\xrightarrow{G} -\frac12(e^{i\phi_1}(a_1^\dagger + a_2^\dagger - a_3^\dagger + a_4^\dagger) -e^{i\phi_2}(a_1^\dagger + a_2^\dagger + a_3^\dagger - a_4^\dagger)) \\ &= -\frac12(e^{i\phi_1} - e^{i\phi_2})(a_1^\dagger + a_2^\dagger) + \frac12 (e^{i\phi_1} + e^{i\phi_2})(a_3^\dagger - a_4^\dagger) \\ &= -C(a_1^\dagger + a_2^\dagger) + B(a_3^\dagger - a_4^\dagger).\label{eq:2}
\end{align}
\end{subequations}
The most recent string of equations show that $(a_3^\dagger - a_4^\dagger)$ maps directly into itself after each round trip. In other words, this superposition of field modes is an eigenmode, or supermode, of the coupled-cavity resonator formed by the Grover-Michelson device. In general, the eigenmodes in these systems can be found by diagonalizing the discrete-time evolution matrix formed from the set of coupled-mode equations of the system.

This recursion can be explicitly unrolled into a geometric series and summed like so
\begin{subequations}
\begin{align}
(a_3^\dagger - a_4^\dagger) &\xrightarrow{N} -C (a_1^\dagger + a_2^\dagger) \sum_{n=0}^N B^{n} + B^{N+1}(a_3^\dagger - a_4^\dagger) \\
&\xrightarrow{N\rightarrow\infty} - C(a_1^\dagger + a_2^\dagger)\sum_{n=0}^\infty B^n  \\
&= \bigg(\frac{C}{B - 1} \bigg )(a_1^\dagger + a_2^\dagger).\label{eq:3}
\end{align} 
\end{subequations}
The amplitudes of a state represented in this basis could have alternatively been converted to the same series. In any case, we now find the transient response of the device and combine it with the above to complete the derivation. The Grover coin maps a photon incident on the first port to
\begin{align*}
    a_1^\dagger|0\rangle &\rightarrow \frac12 (-a_1^\dagger + a_2^\dagger + a_3^\dagger + a_4^\dagger)|0\rangle = (A + \frac12 (a_3^\dagger + a_4^\dagger))|0\rangle.
\end{align*} 
Combining this with the above formulas, we see
\begin{align*}
&\xrightarrow{(\ref{eq:1})} \bigg[A - \frac{B}{2}(a_1^\dagger + a_2^\dagger) + \frac{C}{2}(a_3^\dagger - a_4^\dagger)\bigg ]|0\rangle \\ &\xrightarrow{(\ref{eq:3})}\bigg[A - \frac{B}{2}(a_1^\dagger + a_2^\dagger) + \frac{C}{2}\bigg(\frac{C}{B - 1}\bigg )(a_1^\dagger + a_2^\dagger)\bigg]|0\rangle.
\end{align*}
Grouping by operator yields the output state (\ref{eq:gmi})
\begin{align}
|\psi_{\text{out}}\rangle =\bigg [
\bigg (\frac{C^2}{2B - 2} - \frac{B}{2} - \frac12\bigg )a_1^\dagger + \bigg (\frac{C^2}{2B - 2} - \frac{B}{2} + \frac12\bigg )a_2^\dagger&\bigg]|0\rangle.
\end{align}
Up to an immeasurable global phase shift, the $B$ and $C$ are respectively $r$ and $t$ for the traditional Michelson. Hence the above calculation illustrates that the Grover coin nonlinearly maps the scattering parameters of the standard Michelson interferometer. Because the Grover coin is permutation symmetric there is no need to consider the initial state $a_2^\dagger |0\rangle$. Relabeling ports $1 \longleftrightarrow 2$ results in the same permutation of the output amplitudes so that $r$ and $t$ are the same for input on either port.
\bibliography{refs}
\end{document}